\documentclass[aps,prb,twocolumn,superscriptaddress,showpacs,floatfix,amsmath,amssymb]{revtex4}
\usepackage{graphicx}
\usepackage{color}
\renewcommand{\vec}[1]{ \boldsymbol{#1} }
\renewcommand{\vr}[0]{ \vec{r} }

\newcommand{\nn}[0]{ \nonumber \\}
\newcommand{\lP}[0]{ \left( }
\newcommand{\rP}[0]{ \right) }
\newcommand{\lB}[0]{ \left[ }
\newcommand{\rB}[0]{ \right] }

\newcommand{\lA}[0]{ \left\langle }
\newcommand{\rA}[0]{ \right\rangle }

\newcommand{\Res}[1]{ \mathrm{Res} \lB #1 \rB }
\newcommand{\mat}[1]{ \underline{\vec{#1}} }
\newcommand{\ahalf}[0]{ \frac{1}{2} }

\newcommand{\indll}[3]{ \int_{#2}^{#3}\!\!\mathrm{d}#1 \; }
\newcommand{\dw}[1]{ \int_{-\infty}^{\infty}\!\!\frac{\mathrm{d}#1}{2 \pi} \; }
\newcommand{\dwi}[1]{ \int_{-\infty}^{\infty}\!\!\frac{\mathrm{d}#1}{2 \pi i} \; }
\newcommand{\dz}[2]{ \int_{\!#2}\!\frac{\mathrm{d}#1}{2 \pi} \; }

\newcommand{\FF}[0]{ \hat{\phi}^{\phantom{\dagger}} }
\newcommand{\FFd}[0]{ \hat{\phi}^\dagger }

\newcommand{\eak}[0]{\epsilon^{\phantom{\dagger}}_{\alpha k}}

\newcommand{\eakB}[0]{\bar{\epsilon}^{\phantom{\dagger}}_{\alpha k}}
\newcommand{\Va}[0]{\vec{V}^{\phantom{\star}}_{\alpha}}

\newcommand{\Vak}[0]{\vec{V}^{\phantom{\star}}_{\alpha k}}
\newcommand{\Vakc}[0]{\vec{V}^{\star}_{\alpha k}}

\newcommand{\Vakcp}[0]{\vec{V}^{\star}_{\alpha' k'}}

\newcommand{\kB}[0]{k_\mathrm{B}}

\newcommand{\drv}[0]{\mathrm{drv}}

\newcommand{\Ret}[0]{\mathrm{R}}
\newcommand{\Adv}[0]{\mathrm{A}}
\newcommand{\Mat}[0]{\mathrm{M}}

\newcommand{\Cw}[0]{\mathcal{C}_\omega}

\newcommand{\C}[1]{ \cos \!\lP #1 \rP }
\renewcommand{\S}[1]{ \sin \!\lP #1 \rP }

\begin{document}

\title{Temperature-driven transient charge and heat currents in nanoscale conductors}

\author{F. G. Eich}
\email[]{eichf@missouri.edu}
\affiliation{Department of Physics, University of Missouri-Columbia, Columbia, Missouri 65211}

\author{M. \surname{Di Ventra}}
\affiliation{University of California - San Diego, La Jolla, CA 92093}

\author{G. Vignale}
\affiliation{Department of Physics, University of Missouri-Columbia, Columbia, Missouri 65211}

\date{\today}

\begin{abstract}
  We analyze the short-time behavior of the heat and charge currents through nanoscale conductors
  exposed to a temperature gradient. To this end we employ Luttinger's thermo-mechanical potential
  to simulate a sudden change of temperature at one end of the conductor. We find that the
  direction of the charge current through an impurity is initially opposite to the direction of
  the charge current in the steady-state limit. Furthermore we investigate the transient propagation of
  energy and particle density  driven by a temperature variation through a conducting nanowire.
  Interestingly, we find that the velocity of the wavefronts of, both, the particle and the energy wave
  have the same constant value, insensitive to changes in the average electronic density. 
  In the steady-state regime we find that, at low temperatures, the
  local temperature and potential, as measured by a floating probe lead, exhibit characteristic 
  oscillations due to quantum interference, with a periodicity that corresponds to half the Fermi
  wavelength of the electrons. 
\end{abstract}


\maketitle

\section{ Introduction } \label{SEC:introduction}

The description of the combined charge and energy transport at the nanoscale has
received a great deal of attention in recent years.\cite{NolasGoldsmid:01,DubiDiVentra:11,DiVentra:08}
Much of the motivation is supplied by the search for efficient thermoelectric devices, which would allow, for example, partial conversion of waste heat into usable energy. Experimentally, several procedures have been developed to measure local temperatures  at the nanometer scale, e.g., scanning thermal microscopy \cite{Majumdar:99,YuKim:11,KimLee:11,KimReddy:12,MengesGotsmann:12}
and transmission electron microscopy.\cite{MecklenburgRegan:15}
On the theoretical side various approaches have been used to formally justify
the extrapolation of well-established concepts of equilibrium statistical mechanics, such as temperature and entropy,
to nonequilibrium nanoscale systems.\cite{DubiDiVentra:09,SanchezLopez:13,BergfieldDiVentra:15,BergfieldStafford:13,BieleRubio:15,ShastryStafford:15}

A very interesting theoretical tool for the study of thermoelectric transport phenomena  is the space- and time-dependent thermo-mechanical potential $\psi(\vr,t)$, which was first introduced by Luttinger\cite{Luttinger:64a} to formulate the response of electrons to temperature gradients as a Hamiltonian problem.  Like the gravitational field, to which it is formally related, the thermo-mechanical potential is  linearly coupled to the energy density, for which Luttinger chose one of several possible definitions -- all equivalent in the long-wavelength limit.

In recent years, Luttinger's idea has found several interesting applications in the calculation of
the linear thermoelectric response of macroscopic systems.\cite{Shastry:09,QinShi:11,Shitade:14,Tatara:15a,Tatara:15b}
In a recent paper, we have shown that the thermo-mechanical potential offers a natural path to the
inclusion of thermoelectric effects in a general-purpose time-dependent density-functional theory.\cite{EichVignale:14a} 
Furthermore, we have shown that, when certain dynamical many-body effects are neglected, the thermo-mechanical potential formalism reproduces the results of the well-known Landauer-B\"uttiker~\cite{Landauer:57,BuettikerPinhas:85,Landauer:89} multi-terminal formalism for thermal transport \cite{EichVignale:14b} (see also Refs. \onlinecite{Cini:80,StefanucciAlmbladh:04} for a description of the so-called partition-free approach to quantum transport and its relation to the Landauer-B\"uttiker formalism)  and allows a natural definition of the local temperature in terms of a local probe that carries no currents.\cite{Stafford:14,EichVignale:14b,ShastryStafford:15}

The study of Ref.~\onlinecite{EichVignale:14b} focused on the steady-state response to voltage and temperature gradients. In this paper we present the first application of Luttinger's thermo-mechanical potential approach 
to the computation of transient particle and energy or heat currents through nanoscale devices.

We consider two model devices.  The first one is a single impurity (quantum dot) sandwiched between two thermal reservoirs at different temperatures.  The second model is a conducting chain of atoms placed between the same two reservoirs. 
In both cases we study the time evolution of the electronic and energy densities and the associated currents following a sudden change in the temperature of the left reservoir.  This idealized set up, can actually be approximately realized in experiments, when the current in one of the heater coils typically used to set the temperatures of the reservoirs is suddenly increased or decreased.\cite{DubiDiVentra:11} Of course, in experiments the rate at which the temperature changes in that reservoir is limited by the inelastic processes involved (phonon and radiative dissipation). 
Here, since we are only interested in the quality behavior, we consider an instantaneous switch-on of the perturbation. 
 
In the case of a single impurity, we find that the particle current for short times after the switch-on 
flows opposite to the particle current in the long-time (steady-state) limit,
usually addressed  within the Landauer-B\"uttiker\cite{Landauer:57,BuettikerPinhas:85,Landauer:89}
or Meir-Wingreen approach.\cite{MeirWingreen:92,WingreenMeir:93,JauhoMeir:94}
This remarkable physical effect will be described in detail in the following sections.

Coming to the conducting chain model, we find that the density and energy wavefronts, induced by the sudden change in temperature in the left reservoir, propagate with constant velocity, independent on the initial temperature or density.
In contrast to this,  at low temperatures we find characteristic Friedel oscillations in the steady-state distribution of the local temperature~\cite{DubiDiVentra:09,CasoLozano:10,BergfieldDiVentra:15}, a hallmark sign of quantum interference, with a periodicity depending on the Fermi wavelength, and, hence, the density.

In the present work we have not included the effect of electron-electron interaction. This can be taken into account within the framework of our recently proposed thermal
Density-Functional Theory\cite{EichVignale:14a} with a suitable, e.g., local, approximation for the exchange-correlation potentials.

This paper is organized as follows: In Sec.\ \ref{SEC:ThermoelectricTransport} we introduce the model Hamiltonian employed to study the transient particle and energy density and their associated currents.
In Sec.\ \ref{SEC:Impurity} we present a careful analysis of the transient behavior
of a single-site impurity subject to a temperature gradient. Next, we discuss in Sec.\ \ref{SEC:Wire}
the density and energy wave induced in a nanowire. Details of the numerical implementation are
given in App.\ \ref{APP:Numerics}. We conclude in Sec.\ \ref{SEC:conclusion} by summarizing our findings
and providing an outlook on the implications for thermal Density-Functional Theory.

\section{ Thermoelectric transport in nanoscale junctions } \label{SEC:ThermoelectricTransport}

\begin{figure}
  \includegraphics[width=.48\textwidth]{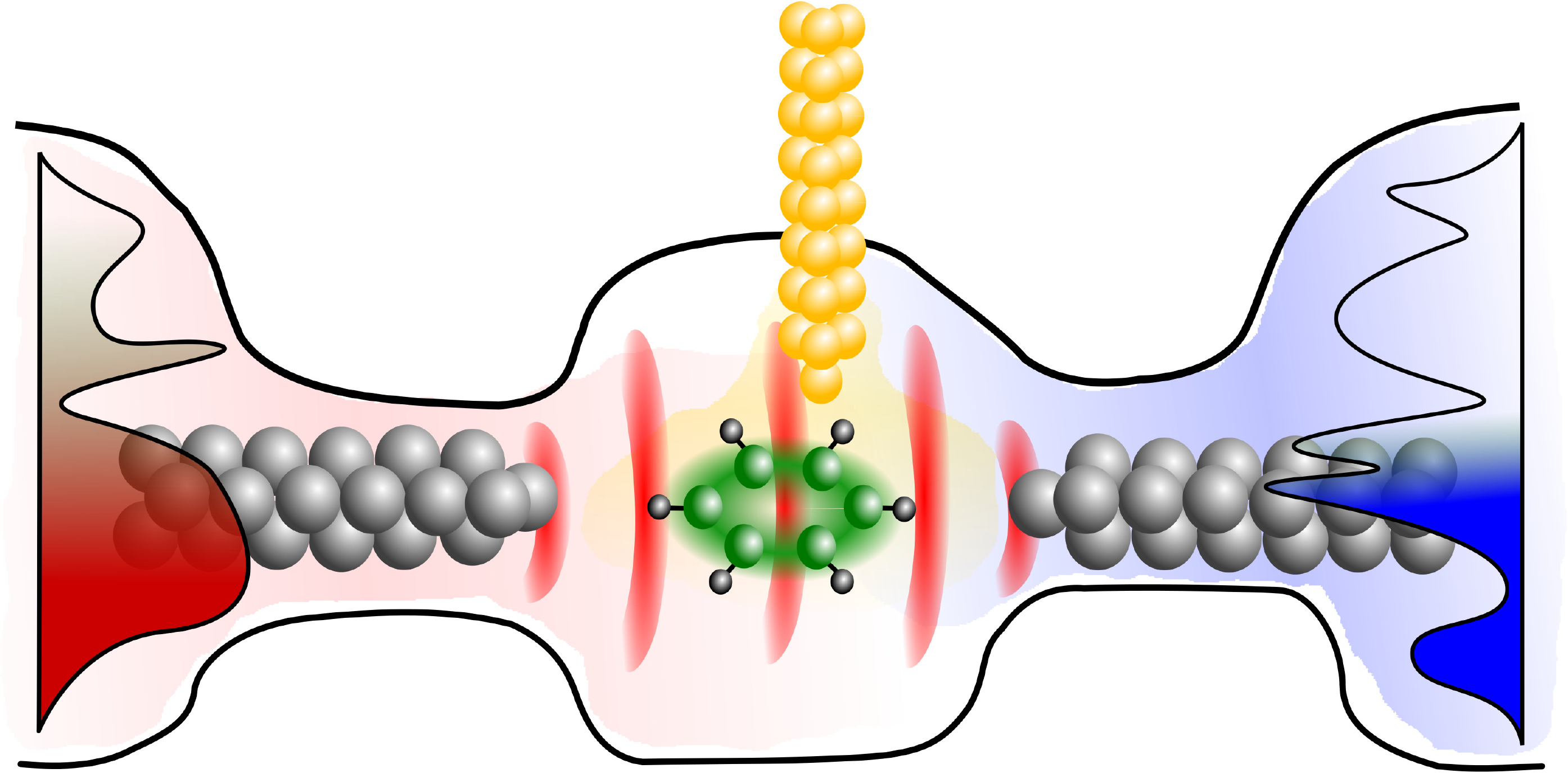}
  \caption{(color online) Schematic  transport setup where a nanoscale junction (central region) is
    connected to reservoirs via leads. When a temperature difference is applied to
    the leads by switching a thermo-mechanical potential a heat and charge current will
    flow through the device. The third lead (yellow) represents an additional ``probe'' lead, which can
    be used to determine the local temperature in the device. The spectral
    densities of the left and right reservoirs are also sketched, as they do, in principle, affect
    the transport.\label{FIG:LeadsJunction}}
\end{figure}

A typical setup to study thermoelectric transport is shown in Fig.\ \ref{FIG:LeadsJunction}, where
a molecular device or a nanowire is suspended between two metallic leads. If the device is
exposed to a temperature gradient, e.g., by heating up the left lead, a heat or energy current
and a charge current flow through the junction.

We model the aforementioned nano-junction by a tight-binding Hamiltonian of the form
\begin{align}
  \hat{\mathcal{H}} & = \sum_{\alpha k} \eak \FFd_{\alpha k} \FF_{\alpha k} + \sum_{mn} \FFd_n H_{nm} \FF_m \nn
  & \phantom{=} {} + \sum_{\alpha k} \sum_m \lP \FFd_{\alpha k} V_{(\alpha k) m} \FF_m 
  + \FFd_m V_{m (\alpha k)} \FF_{\alpha k} \rP ~. \label{Hamiltonian}
\end{align}
where ${\alpha}$ labels leads connected to the central region. The electrons in the leads are
governed by a dispersion ${\eak}$. We model the leads by an infinite tight-binding chain
with nearest-neighbor hopping amplitudes $t_\alpha$, which means that the dispersion
reads explicitly
\begin{align}
  \eak & = 2 t_\alpha \C{k} + \epsilon_\alpha ~. \label{leadDispersion}
\end{align}
It describes a single band with bandwidth $4 t_\alpha$ and
the positioning of the center of the band is determined by the lead-specific energy $\epsilon_\alpha$.

The central region is described by the generic Hamiltonian ${\mat{H}}$,
a matrix in the basis of the tight-binding sites which contains the kinetic energy, 
described by a uniform nearest-neighbor hopping $t$, and a local potential $U_n$, i.e.,  
\begin{align}
  H_{mn} & = t \delta_{m (n \pm 1)} + U_n \delta_{mn} ~, \label{ImpurityHamiltonian}
\end{align}

The hopping amplitudes between leads and impurity are denoted by ${\Vak}$.
Taking only a nearest-neighbor hopping between the last lead site
and the closest site of the central region with amplitude $\Va$ we have
\begin{align}
  \Vak & = \Va \S{k} ~. \label{hoppingAmplitudes}
\end{align}
In the limit of infinite leads the sum over $k$
corresponds to $\sum_k \equiv \frac{2}{\pi} \indll{k}{0}{\pi}$.
Hamiltonian \eqref{Hamiltonian} describes the intrinsic features
of the system under consideration.

Usually, temperature-driven transport is described by removing the contacts between
the central region and the leads in the initial preparation, and equilibrating the
leads at different temperatures.\cite{DiVentra:08,ToppSchaller:15}
Then, at the initial time $t_0$, the device is suddenly contacted to the leads which
induces a heat and charge transfer trough the central region. 
Here, by contrast, the initial state is determined for the fully contacted system.
This is possible since we are employing Luttinger's thermo-mechanical potential
to describe a gradient in the temperature. At
$t_0$ we switch on a \emph{thermal} and \emph{charge} bias in the leads.
This means that the Hamiltonian for $t>t_0$, which drives the system out
of equilibrium is given by
\begin{align}
  \hat{\mathcal{H}}_\drv & = \sum_{\alpha k} \eakB \FFd_{\alpha k} \FF_{\alpha k}
  + \sum_{mn} \FFd_n H_{nm} \FF_m \nn
  & \phantom{=} {} + \sum_{\alpha k} \sum_m \lP \FFd_{\alpha k} V_{(\alpha k) m} \FF_m 
  + \FFd_m V_{m (\alpha k)} \FF_{\alpha k} \rP ~, \label{DrivingHamiltonian}
\end{align}
where the dispersion in the leads has changed to
\begin{align}
  \eak \to \eakB = (1 + \psi_\alpha) (\eak + U_\alpha) ~. \label{DrivingDispersion}
\end{align}
The potential bias $U_\alpha$ shifts the center of the band and the thermal bias $\psi_\alpha$
stretches the shifted bands. We have shown in a previous work,\cite{EichVignale:14b} that the application
of the thermal bias $\psi_\alpha$ corresponds to changing the temperature in lead $\alpha$
by $\delta T_\alpha = \psi_\alpha T_0$, i.e., $\psi_\alpha$ determines the relative temperature change.

\section{Transient currents for a single-site impurity } \label{SEC:Impurity}

\begin{figure}[t]
  \includegraphics[width=.48\textwidth,clip]{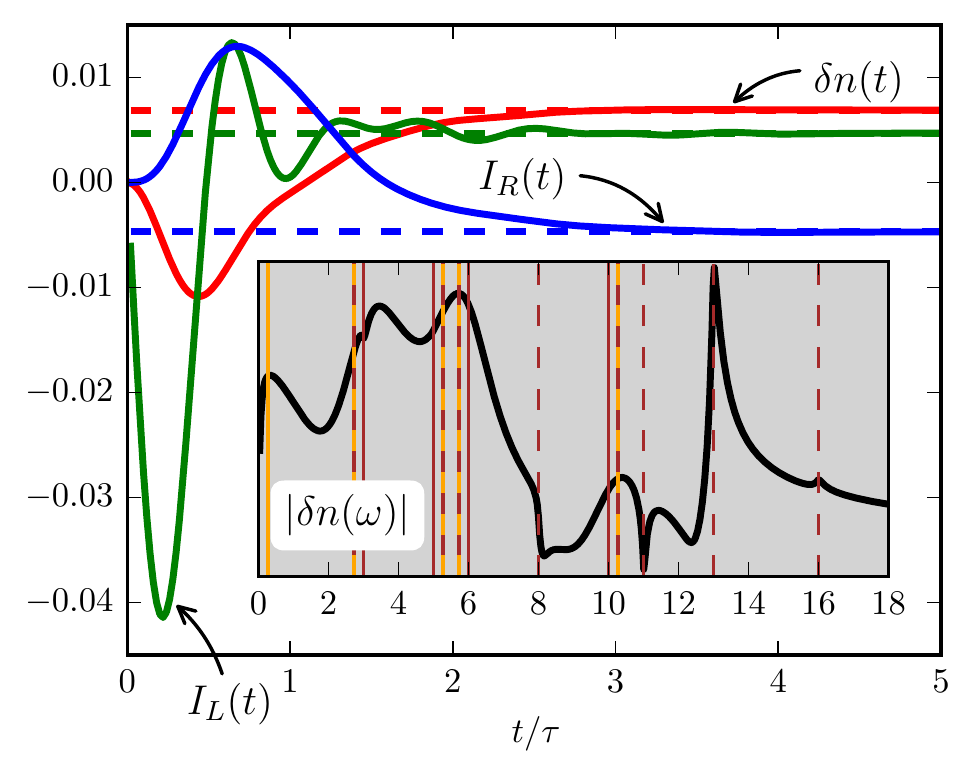}
  \caption{(color online) Plot showing the transient density change, $\delta n(t)$,
  of the impurity site and the currents between left lead and impurity, $I_L(t)$, and right lead
  and impurity,$I_R(t)$, respectively. The corresponding steady-state value are indicated
  by the horizontal, dashed lines. The inset shows the Fourier transform of the density change
  of the long-time tail. The structure of this power spectrum reflects the distribution of energy levels in the leads.\label{FIG:OneSiteTDDensity}}
\end{figure}
As a first example we consider a single-site impurity (quantum dot) coupled to two
(symmetric) metallic leads. Specifically, we take the impurity site to be 
aligned with the chemical potential and the hopping amplitudes between the impurity and the
leads are chosen as our unit of energy, i.e.,  $V_\alpha=V=1$.(See also Appendix~\ref{APP:Numerics} for more details.) 

The hopping amplitudes in both leads is $t_\alpha = 2 V$,
which means that the leads have a bandwidth of $8 V$. Both leads are shifted down
in energy by $-1 V$ in order to break particle--hole symmetry, which is required
to observe the Seebeck or Peltier effect, i.e., the interplay between charge and energy.\cite{DubiDiVentra:11} Accordingly, both the left and right lead have
band edges which are positioned at $-5V$ (lower band edge) and $+3V$ (upper band edge)
measured from the chemical potential, which is taken to define zero energy.

We stress that we do not take the wide-band limit. Accordingly, the embedding self-energy
due to the leads does not only provide a finite lifetime for the impurity state, but also
shifts its energy. For leads modeled by a tight-binding chain this shift is linear--as long as the
impurity site lies within the band--and pushes the energy of the impurity
above the chemical potential in the present scenario.

Initially the coupled system is equilibrated at a temperature $\kB T_0 = 0.25 V$. Then, at $t=0$,
the temperature in the left lead is suddenly raised by applying a thermo-mechanical
potential $\psi_L=1$, which corresponds to a doubling of the temperature on the left
side.

In Fig.\ \ref{FIG:OneSiteTDDensity} we show the transient change of the impurity density
and the currents flowing to the left and right lead, respectively.
A temperature-driven particle current occurs only because
the system is not particle--hole symmetric. A perfect alignment of the center of both bands
with the chemical potential and the impurity site would have two effects:
1) The energy of the impurity state would not be shifted, because the real part of the 
embedding self-energy vanishes at the center of the band. 2) The transmission
would be symmetric, which implies that no net particle current flows.

The time scale $\tau$ in the plot of the transients in Fig.\ \ref{FIG:OneSiteTDDensity}
represents the intrinsic time scale for the decay of electrons into the leads.
The embedding self-energy due to lead $\alpha$ is proportional to $V_\alpha^2 /t_\alpha$, which, in turn, implies a that the lifetime of the electrons
due to the embedding is $\tau_{\alpha} \propto \hbar t_\alpha / 2 V_\alpha^2$. Since there are two
leads we add the decay \emph{rates} to get $\tau = \hbar t_\alpha / (2 V^2) = \hbar / V$.
For times $t>\tau$ the density (red line) and the
currents from the left lead (green line) and the right lead (blue line) approach their respectively
steady-state values (dashed lines). As expected, the current from the right lead is the negative of the current from the
left lead in the steady-state regime. Furthermore the density change settles to a positive value which
means that in the transient regime the impurity acquires additional particles. This can be expected since the
impurity will increase its temperature due to the heating from the left lead. We recall that the
energy of the impurity site is above the chemical potential, due to the coupling to the metallic leads,
and hence a higher temperature results in an increase in density.

\begin{figure}[h]
  \includegraphics[width=.48\textwidth,clip]{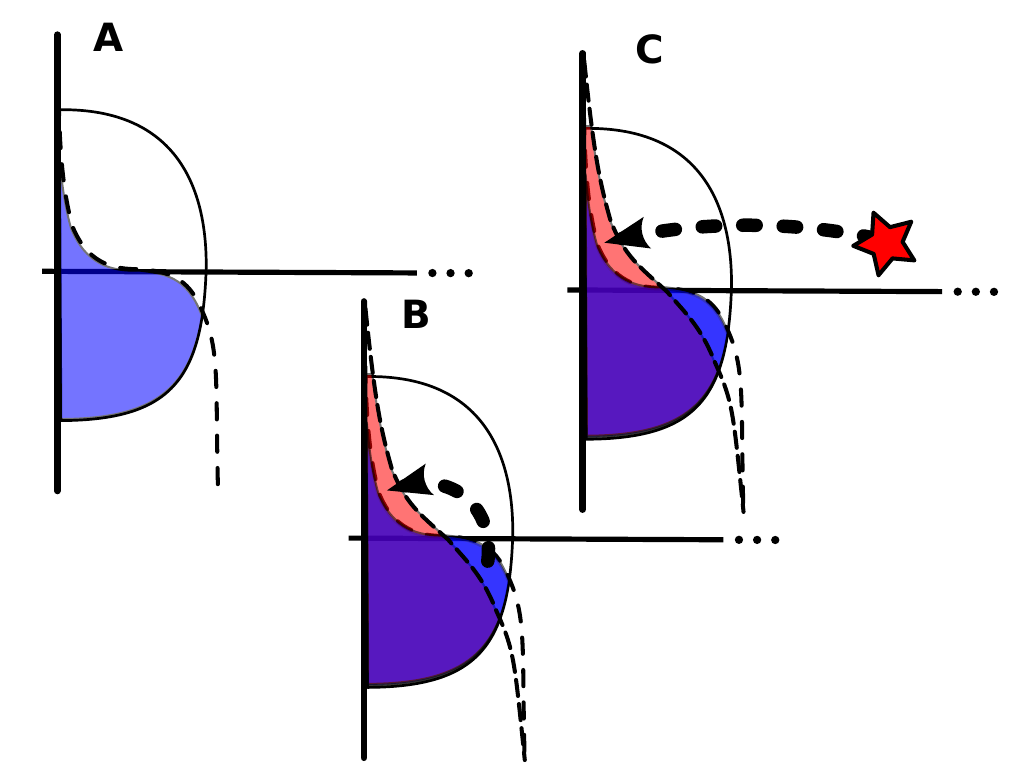}
  \caption{(color online) Sketch showing the short time particle transfer
  processes: (A) The initial occupation of the left lead. (B) The sudden increase in temperature
  requires a redistribution of the electrons from below to above the chemical potential. (C)
  The presence of the impurity assist--at short times--this redistribution by providing
  electrons above the chemical potential and the impurity density drops initially. \label{FIG:OneSiteImpuritySketch}}
\end{figure}
Turning to the short-time transient, i.e., $t < \tau$, we see that the density of the impurity
decreases, which seems to be counterintuitive. However we suggest a simple picture
(cf.\ Fig.\ \ref{FIG:OneSiteImpuritySketch}):
The thermo-mechanical potential applied to the left lead forces the electrons to adjust to a higher temperature.
This means that electrons have to be moved from below the chemical potential to above the chemical potential.
The presence of the impurity site can facilitate this process, at least 
temporarily, by providing electrons above the chemical potential. This means that for short times
electrons are "sucked" into the left lead, which decreases the impurity density. However,
the impurity will have to take a higher temperature, and by extension density, itself. Now
the right lead comes into play by providing electrons for the impurity.

This explanation is supported by the analysis of the transient currents. Initially there is a very strong flow from the impurity to the left lead
($t<0.5\tau$). A little later we observe a flow from the right lead to the impurity $0.25 \tau  < t < \tau$.
Finally, the two currents cross and settle at opposite steady-state values.

\begin{figure}
  \includegraphics[width=.48\textwidth]{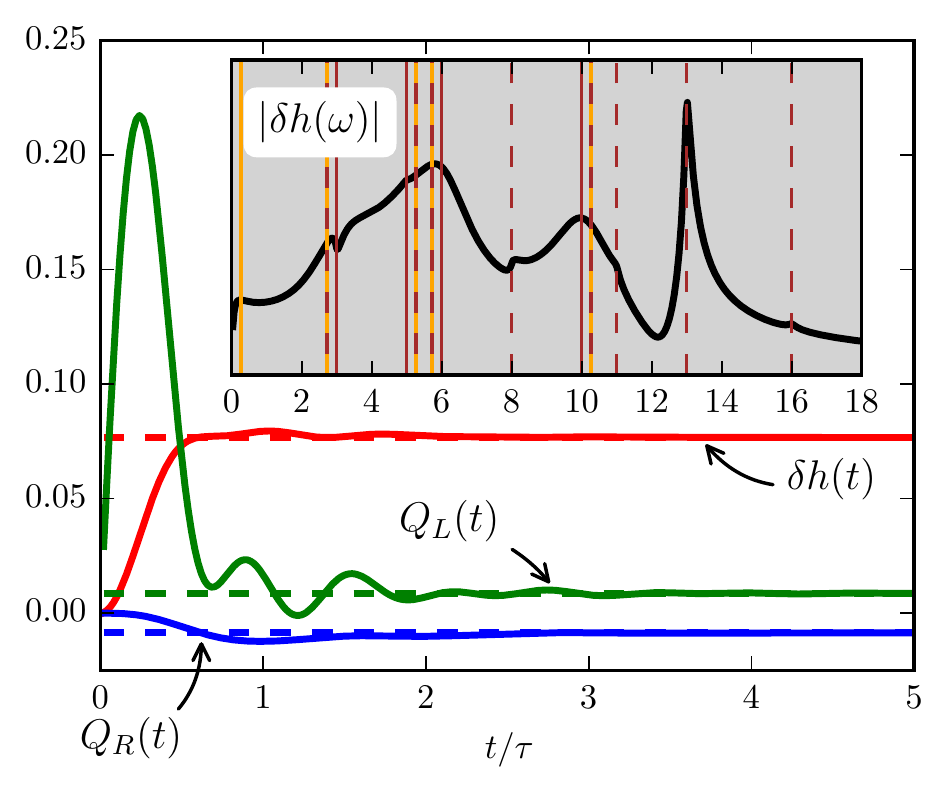}
  \caption{(color online) Same as Fig.\ \ref{FIG:OneSiteTDDensity}
    but for the change in impurity energy, $\delta h(t)$, and left and right
    heat currents, $Q_L(t)$ and $Q_R(t)$, respectively. \label{FIG:OneSiteTDEnergy}}
\end{figure}
In Fig.\ \ref{FIG:OneSiteTDEnergy} we show the time evolution of the impurity energy (red line)
and its associated heat currents from the left lead (green line), and from the right lead (blue line). Since we heat
up the system, it is always expected that the impurity energy increases, independent of the positioning
of the impurity level. This is simply due to the fact that the energy is measured with respect to the
chemical potential. Even if the impurity level would be below the chemical potential, which means that the
state depopulates in the steady state, the change in energy would be positive,
because we depopulate a \emph{negative} energy state.

In light of the previous discussion of the particle flow at short times one may ask how it is possible
to have a heat flow from the left lead to the impurity even though there are electrons
moving above the chemical potential in the opposite direction. The resolution to this puzzle is the following:
The energy of the impurity site is given by the impurity density times the local potential plus a contribution
due to the hopping between the impurity and both leads.\footnote{We have adapted the convention to
split the hopping energy equally between the participating sites.} In the present case the local potential
is perfectly aligned with the chemical potential, i.e., the contribution from the local potential is zero. Accordingly, the only contribution to the local energy comes from the hopping to the leads.
This (kinetic) energy does not depend
on the ``direction'' of the hopping and therefore the local energy increases. Looking at the heat currents
we see that there is initially a strong heat flow from the left lead to the impurity, followed
by a much less pronounced heat flow from the impurity to the right lead. Finally, the flows equilibrate
to the steady-state values.

The insets in both Figs.\ \ref{FIG:OneSiteTDDensity} and \ref{FIG:OneSiteTDEnergy} show the
Fourier transform of the density and energy change at long times. It is computed
in a time window $t = [10 \tau, 10 \tau + \Delta t]$, where $\Delta t$ is chosen big enough to resolve
the ``lowest'' transition energy of our system, which in our example corresponds to transitions
between the impurity level, $\epsilon_0$, and the chemical potential
($\hbar \omega_{min} = \epsilon_0 \approx 0.27V$,
vertical orange line).
The sampling rate is taken to resolve the largest transition frequency, which is given by the energy
differences of the thermally biased band edges of the left lead ($\hbar \omega_{max} = 16V$).
In order to understand the possible transitions we recall that the band edges are initially at $-5V$
and $+3V$ for both the left and right lead. Applying the thermo-mechanical potential scales the
left band by a factor of $2$, which shifts the band edges of the left lead to $-10V$ and $+6V$, respectively.
The solid, brown vertical lines depict transition frequencies from the band edges
to the chemical potential, i.e., they are at $\hbar \omega = 3V, 5V, 6V, 10V$.
Similarly, the dashed, brown vertical lines highlight transitions between band edges which correspond to
$\hbar \omega = 3V, 5V, 8V, 11V, 13V, 16V$. Lastly, the
dashed, brown--orange vertical lines indicate transitions between the band edges and the impurity level
at $\hbar \omega = 3V - \epsilon_0, 5V + \epsilon_0, 6V - \epsilon_0, 10V + \epsilon_0$.
Strong features of the Fourier spectrum coincide with the aforementioned transition frequencies.
Note that in the wide band limit all features, except for the transition between the impurity level
and the chemical potential at $\hbar \omega = \epsilon_0$, would be absent.
The most distinct peak occurs for, both, the
density and the energy at $\hbar \omega=13V$, which refers to transitions between the
lower band edge of the left lead and the upper band edge of the right lead.

\section{ Heat wave propagation through a conducting wire} \label{SEC:Wire}

Our second example describes a nanowire suspended between two metallic leads.
The parameters for this system are taken to be identical to the single-impurity
model discussed in the previous section. However, the central region is composed of $100$ sites connected by nearest neighbor hopping
with amplitudes $t=V$.
This means that the central region starts to form a band with
bandwidth $4V$ and a dispersion given by the discretized version of Eq.\ \eqref{leadDispersion}.
The center of the band representing the nanowire is aligned with the chemical potential. 

\begin{figure}
  \includegraphics[width=.45\textwidth,clip]{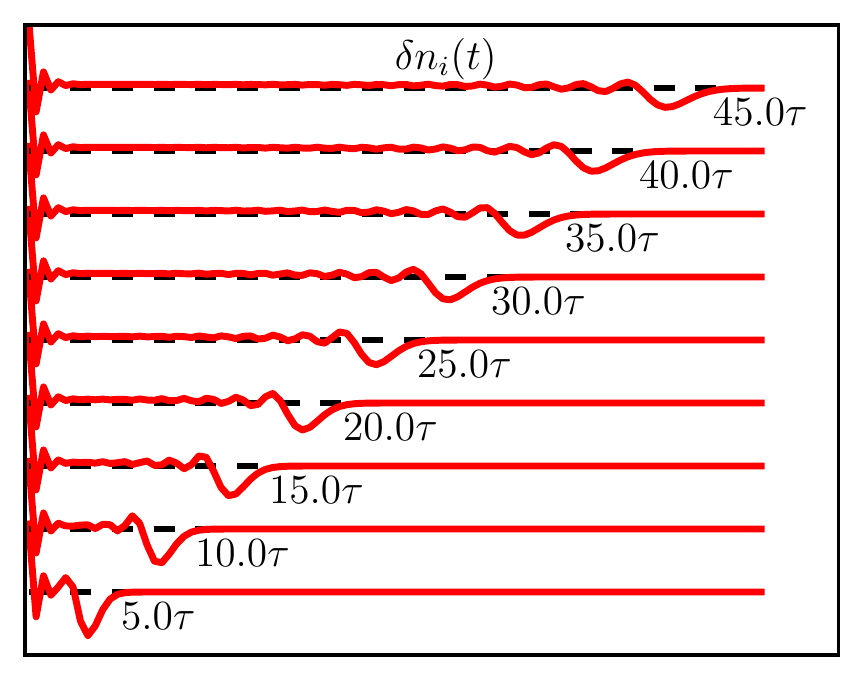}
  \includegraphics[width=.45\textwidth,clip]{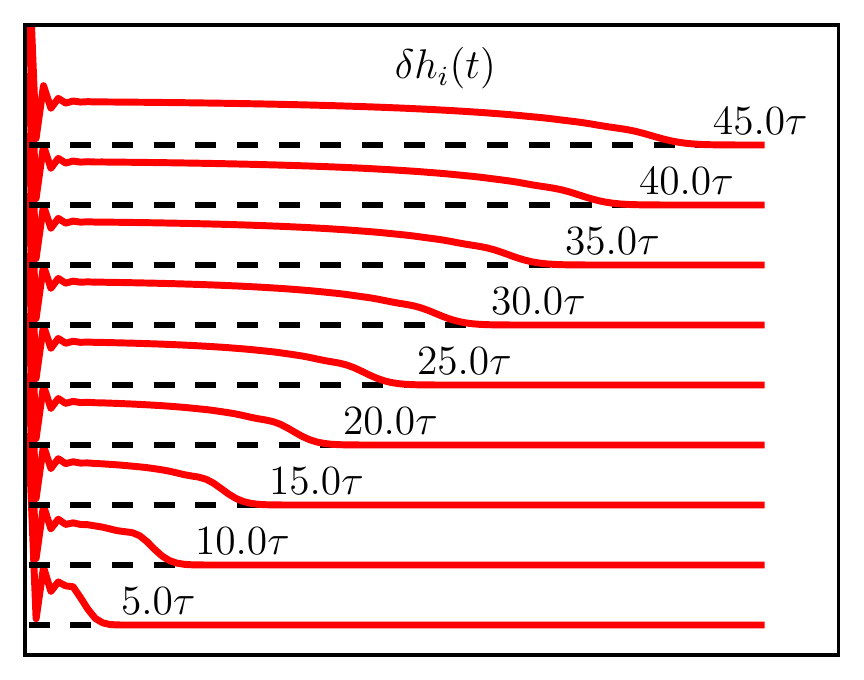}
  \caption{(color online) Plot of the transient density and energy wave
  propagating through the nanowire. No external gate potential is applied.
  Both wavefronts propagate with the same velocity from the left to the right lead.\label{FIG:NanoWire1}}
\end{figure}
In Fig.\ \ref{FIG:NanoWire1} we show snapshots of the spatially-resolved density and energy in the wire.
The snapshots are taken at intervals of $\delta t = 5 \tau$ up to the time $t < 50 \tau$, just
before the wavefronts reach the right end of the wire. First of all, we note that both the density
and the energy wavefronts traverse the wire with the same constant velocity. This ``Wiedemann-Franz''--like
behavior can be understood from the fact that the energy is carried by the propagating electrons. Their
spatial behavior, however, is different in the wake of the wavefront. As expected, the velocity of the wavefront
is proportional to the hopping amplitude, i.e., $v \propto t$.

\begin{figure}
  \includegraphics[width=.45\textwidth,clip]{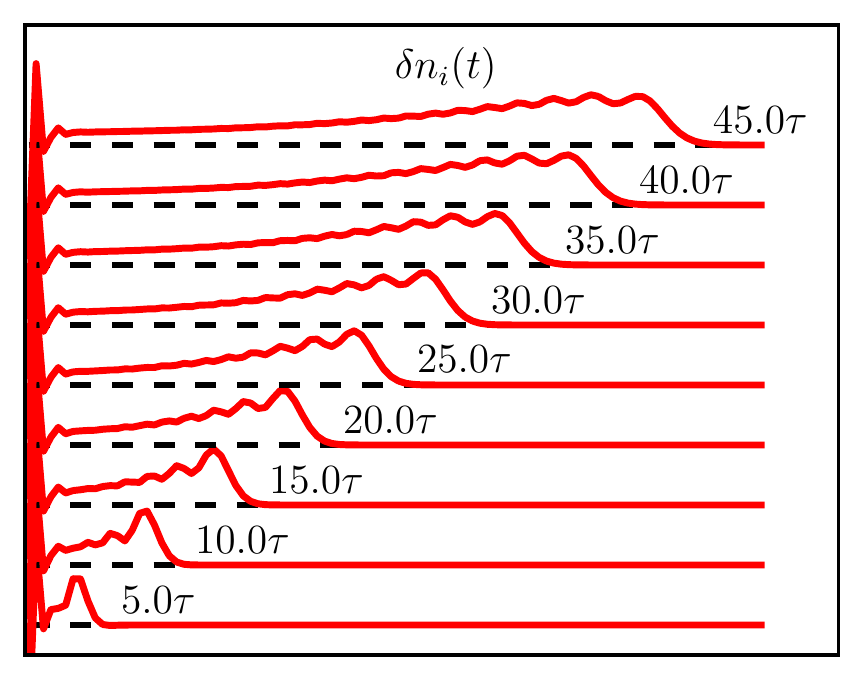}
  \includegraphics[width=.45\textwidth,clip]{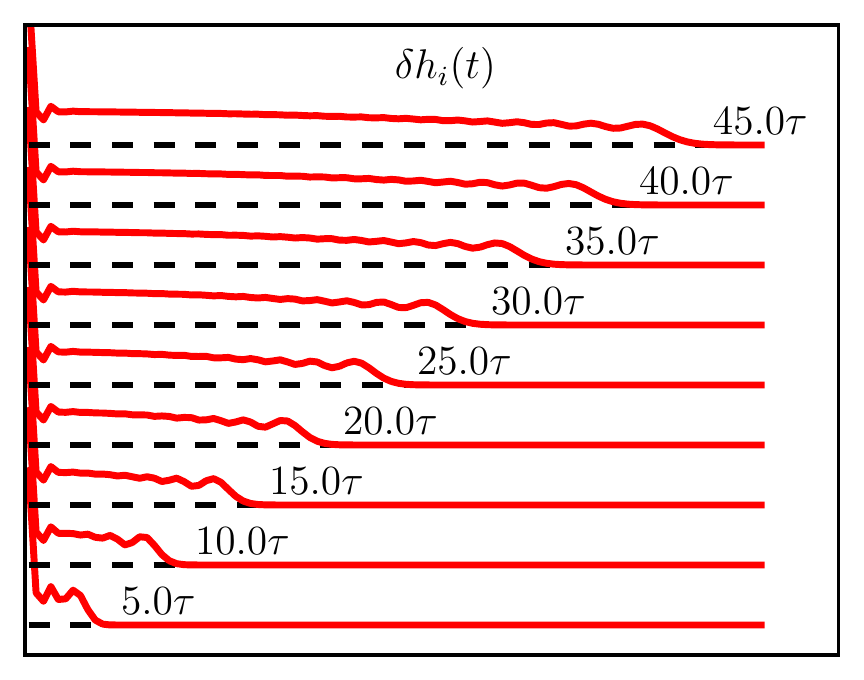}
  \caption{(color online) Same as Fig.\ \ref{FIG:NanoWire1} but with an external
  gate potential. This reduces the initial density--and thereby the Fermi wave vector--of the nanowire.
  The transient wavefronts of the density and energy, however, propagate with the same velocity
  as with no external gate potential. \label{FIG:NanoWire2}}
\end{figure}
A slightly more refined guess for the velocity is the Fermi velocity, $v_\mathrm{F} \propto \partial_k \epsilon_k = 2 t \S{k_\mathrm{F}}$.
This implies a density dependence of the velocity via the Fermi wave vector $k_\mathrm{F}$. In order to investigate whether
there is a density dependence of the velocity we repeat the calculation with the dispersion of the nanowire shifted upwards
by a constant gate potential $U_n=1V$. In Fig.\ \ref{FIG:NanoWire2} we show snapshots of the density and energy changes
for the gated nanowire. While the spatial form of the waves changes compared to the nanowire without any gate potential,
the wavefront still moves with the same velocity. We do not find a density dependence.

Of course, the simplistic estimate of the velocity by the $v_\mathrm{F}$ has two caveats: 1) We inject
a highly inhomogeneous wave packet in the nanowire, which implies that we have a superposition of many
momentum states. Accordingly, it seems rather optimistic to assume that the wave packet is highly peaked
around the Fermi wave vector. 2) The \emph{initial} temperature is comparable to the bandwidth of the nanowire, i.e.,
$\kB T_0 \lesssim 4 t$. Hence, the thermal spread of occupations is of the order of the Fermi energy, 
$\epsilon_\mathrm{F} = -2V \C{k_\mathrm{F}} + 2V$. We have computed the transients of the density and energy 
with an initial temperature reduced by a factor of $10$, i.e., $\kB T_0 = 0.025$. However, we find that--
with and without the gate potential--the velocity of the wavefront corresponds to the velocity
at the higher initial temperature.
This leads to the conclusion that the spatial inhomogeneity of the wavefront requires a superposition of momentum states. We point out that it has recently been shown that the coordination of the tight-binding model affects the velocity of the wavefront.\cite{MetcalfChien:16} It would be interesting to investigate if this geometric effect allows for different propagation velocities for density and energy waves.

\begin{figure}
  \includegraphics[width=.48\textwidth,clip]{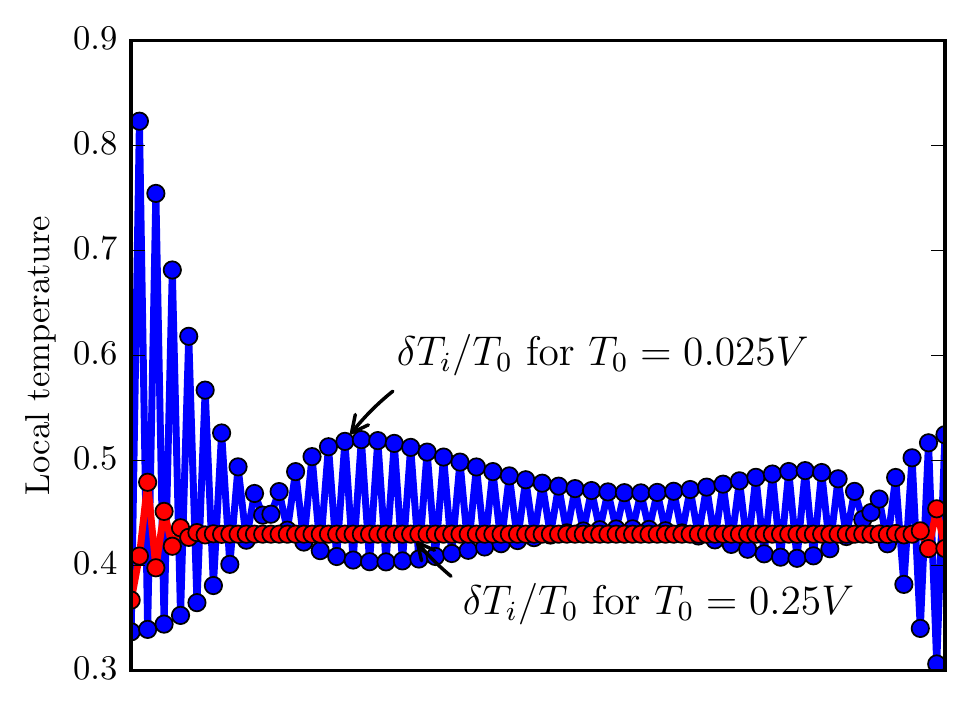}
  \includegraphics[width=.48\textwidth,clip]{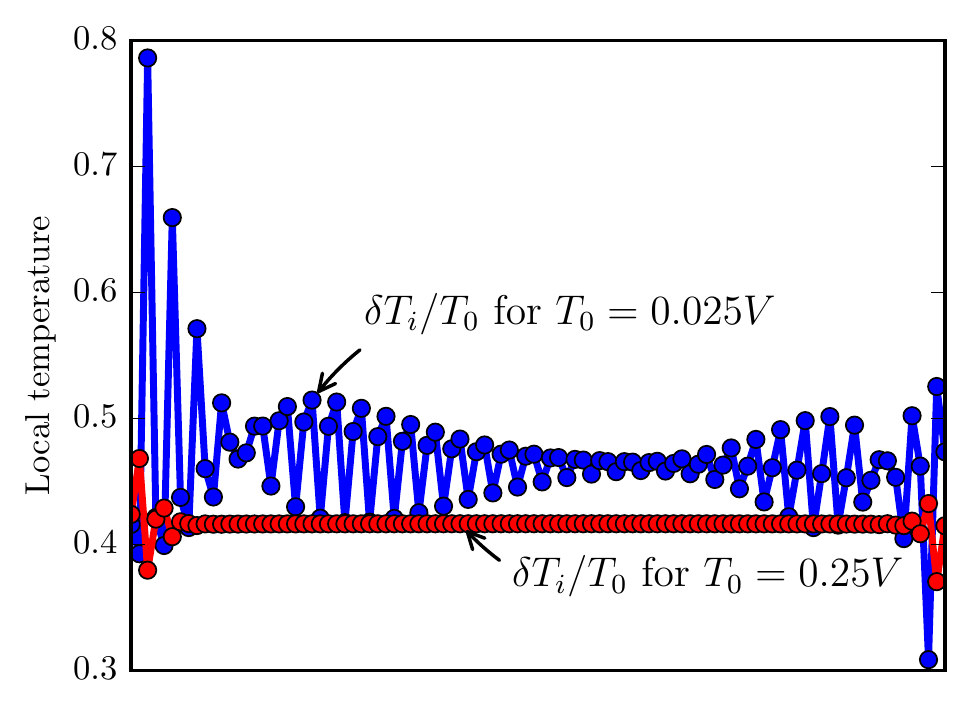}
  \caption{(color online) Steady-state local temperature distributions in the nanowire determined by a
  ``probe'' lead as shown in Fig.\ \ref{FIG:LeadsJunction}. Both panels
  compare the local temperatures for different initial temperatures. The
  red lines correspond to an initial temperature of $\kB T_0 = 0.25V$ and the
  blue lines to $\kB T_0 = 0.025V$. The local temperature for the lower
  initial temperature exhibit typical $2 k_\mathrm{F}$ Friedel oscillations. \label{FIG:NanoWireLocalTemperature}}
\end{figure}
Lastly, we look at the steady-state of the nanowire. We can determine the local temperature and potential
by introducing a third lead (cf.\ Fig.\ \ref{FIG:LeadsJunction}),
which is weakly coupled to a specific site in the wire. Furthermore we take the
wide-band limit for this additional lead.
A local potential and temperature can be defined by imposing zero particle and energy current conditions for this ``probe'' lead.\cite{Stafford:14} 
It has been pointed out by us (cf.\ Ref.\ \onlinecite{EichVignale:14b}) that the zero current conditions are equivalent to
asking: Which temperature and chemical potential reproduce the local density and energy under equilibrium conditions? It was also shown recently~\cite{YeDiVentra:15} that the local 
temperature obtained this way is comparable to that experimentally measurable in which one varies the temperature of the third lead till 
some observable of the system is minimally perturbed.\cite{DubiDiVentra:09} 

In Fig.\ \ref{FIG:NanoWireLocalTemperature} we compare the local temperature computed for different initial
temperatures. The upper panel depicts the local temperature for the wire without the gate. We can see that at
low initial temperature ($\kB T_0 = 0.025V$, blue line) the local temperature oscillates from site to site, whereas
for high initial temperature ($\kB T_0 = 0.25V$, red line) the spatial temperature profile is essentially flat.
The lower panel shows temperature profiles for the gated nanowire. Qualitatively we see the same behavior as for the
wire with no gate potential. However, the oscillations for low initial temperature now have a period of three lattice sites.
The applied gate reduces the Fermi wave vector from
$k_\mathrm{F} = \pi / 2 a_0 \to k_\mathrm{F} = \pi / 3 a_0$ ($a_0$ being the distance between
neighboring sites). Accordingly,
the oscillations in the local temperature correspond in both cases to ``Friedel''--like oscillations at $q=2k_\mathrm{F}$.
Friedel oscillations are a well-known feature of the degenerate electrons gas and represent a quantum interference
effect. The average temperature variation of the wire is slightly below $\delta T / T_0 = 0.5$, i.e.,
the wire is closer in temperature to the colder right lead. We have already observed this phenomenon in Ref.\ \onlinecite{EichVignale:14b},
which was also predicted in Ref.\ \onlinecite{DubiDiVentra:09,BergfieldDiVentra:15}. We conclude by mentioning that the local potential exhibits
the same oscillations. The interested reader may find the corresponding plots in App.\ \ref{APP:additionalPlots}.

\section{ Discussion and conclusion } \label{SEC:conclusion}

In this paper we have investigated the transient currents induced by a temperature gradient.
The temperature gradient has been applied by employing Luttinger's thermo-mechanical potential as proxy for
temperature variations. Furthermore, the formulation in terms of the thermo-mechanical potential allowed us
to study temperature-driven particle and energy transport in the so-called unpartitioned approach, where
a nano scale device is already contacted to metallic leads in the initial preparation.

For a single-site impurity model we found that the transient particle current flows in the opposite direction
to the steady-state current, which suggests that a frequency dependent generalization of the Seebeck coefficient
changes sign at high frequencies. Furthermore, we provided a simple picture to interpret the numerical results
for the transient particle current in terms of a impurity assisted re--population of the electrons in the
leads.

Considering a tight-binding chain, representing, e.g., conductive polymers or nanowires, we found that the velocity of the
transient particle and energy wave is essentially constant over a range of initial temperatures and only depends
on the hopping amplitudes. Furthermore we have shown that in the steady state there is a signature of quantum
interference--at least at low temperatures. The local temperature and potential, as measured by a floating thermal probe
exhibits characteristic $2 k_\mathrm{F}$ Friedel oscillations.

Even though the model studied considered noninteracting particle, the results are highly relevant, since we have recently introduced
a thermal Density-Functional Theory,\cite{EichVignale:14a} which allows to map the interacting system onto a fictitious non-interacting 
Kohn-Sham system.\cite{KohnSham:65} In the future it will be interesting to investigate to what extent
interactions, represented in terms of exchange-correlation corrections to the thermo-mechanical and 
charge potential will affect the presented results. 
We are confident that the presented results are an important step on the way to a fully microscopic  
description of the combined particle and energy transport in interacting systems.

\begin{acknowledgments}
  We gratefully acknowledge support from the Deutsche
  Forschungsgemeinschaft under DFG Grant No. EI 1014/1-1 (F. G. E.), and the
  DOE under Grants No. DE-FG02-05ER46203 (G. V.) and DE-FG02-05ER46204 (M. D.).
\end{acknowledgments}

\begin{appendix}
  \begin{figure}
    \includegraphics[width=.4\textwidth,clip]{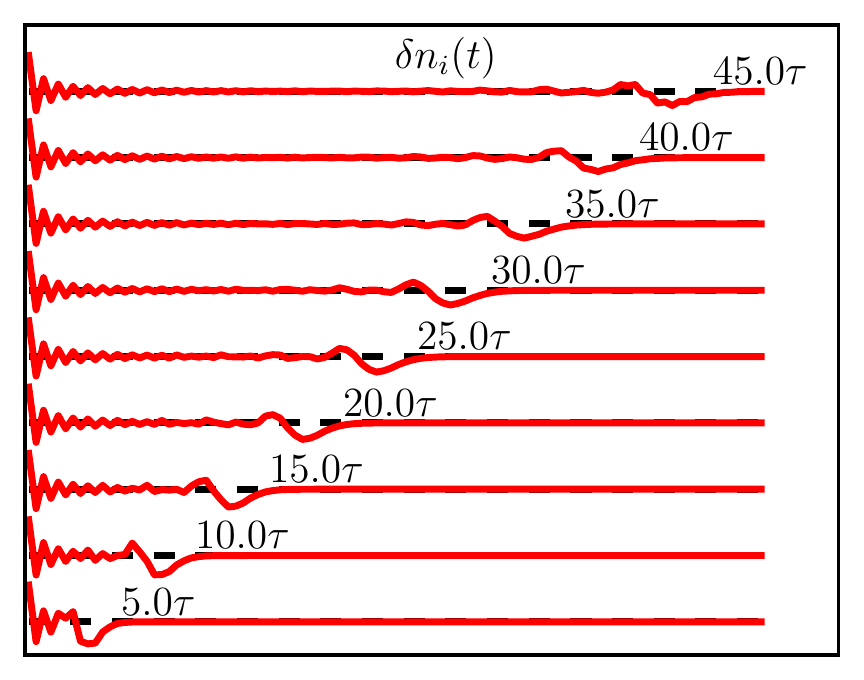}
    \includegraphics[width=.4\textwidth,clip]{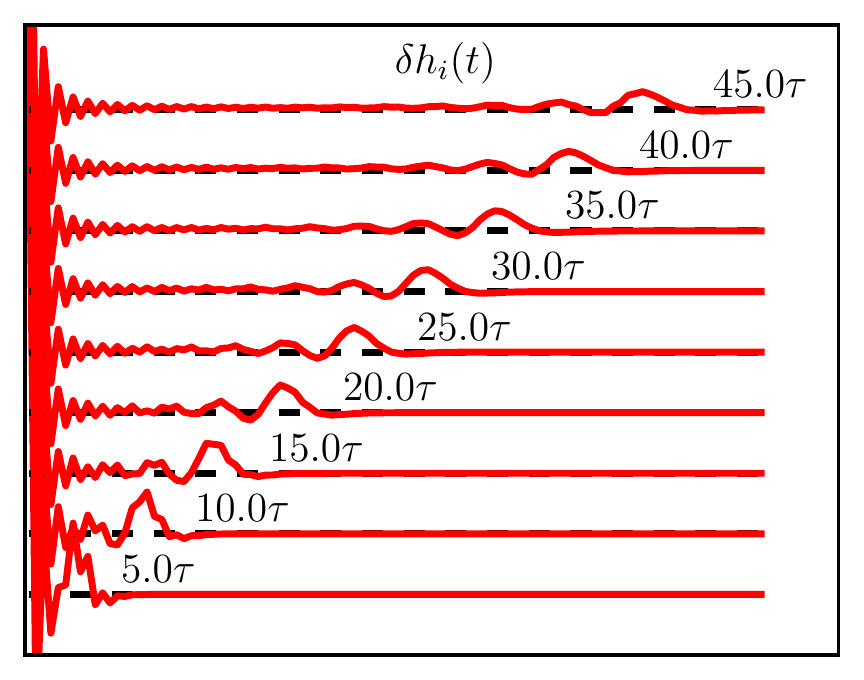}
    \caption{(color online) Same as Fig.\ \ref{FIG:NanoWire1} in Sec.\ \ref{SEC:Wire}
    but with a reduced initial temperature $\kB T_0 = 0.025V$. The velocity
  of the transient wavefront remains unaffected.\label{FIG:NanoWire3}}
  \end{figure}
  \section{Additional plots} \label{APP:additionalPlots}
  \begin{figure}
    \includegraphics[width=.4\textwidth,clip]{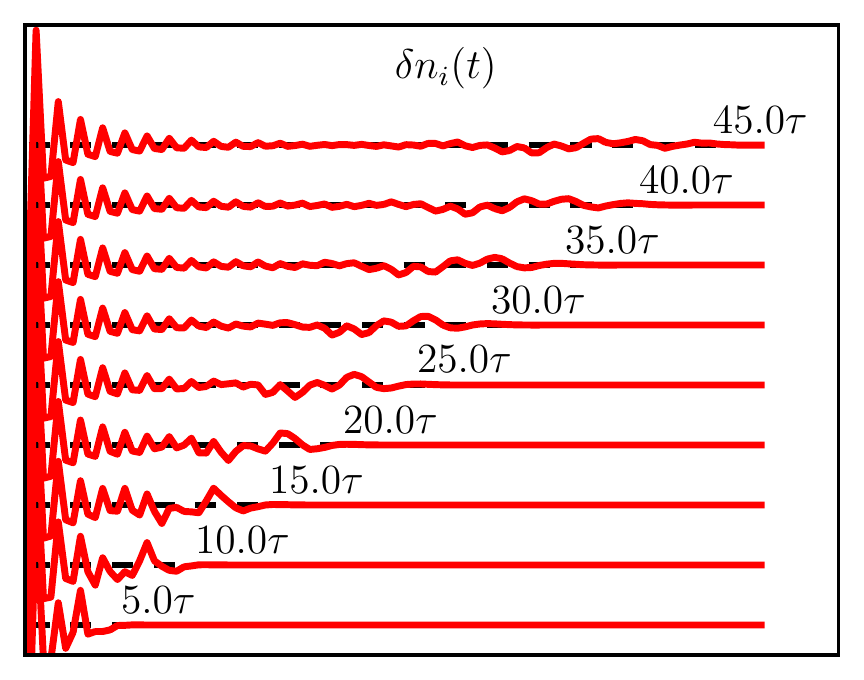}
    \includegraphics[width=.4\textwidth,clip]{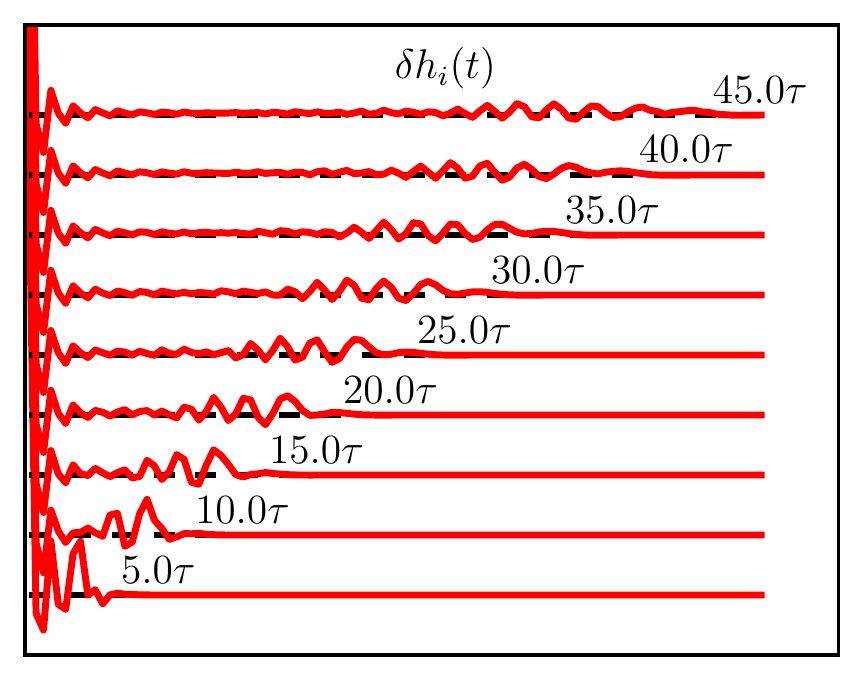}
    \caption{(color online) Same as Fig.\ \ref{FIG:NanoWire2}, Sec.\ \ref{SEC:Wire} but
    with initial temperature  $\kB T_0 = 0.025V$. Also for the case with an externally
    applied gate voltage the velocity of the wavefront is the same.\label{FIG:NanoWire4}}
  \end{figure}
\begin{figure}
    \includegraphics[width=.45\textwidth,clip]{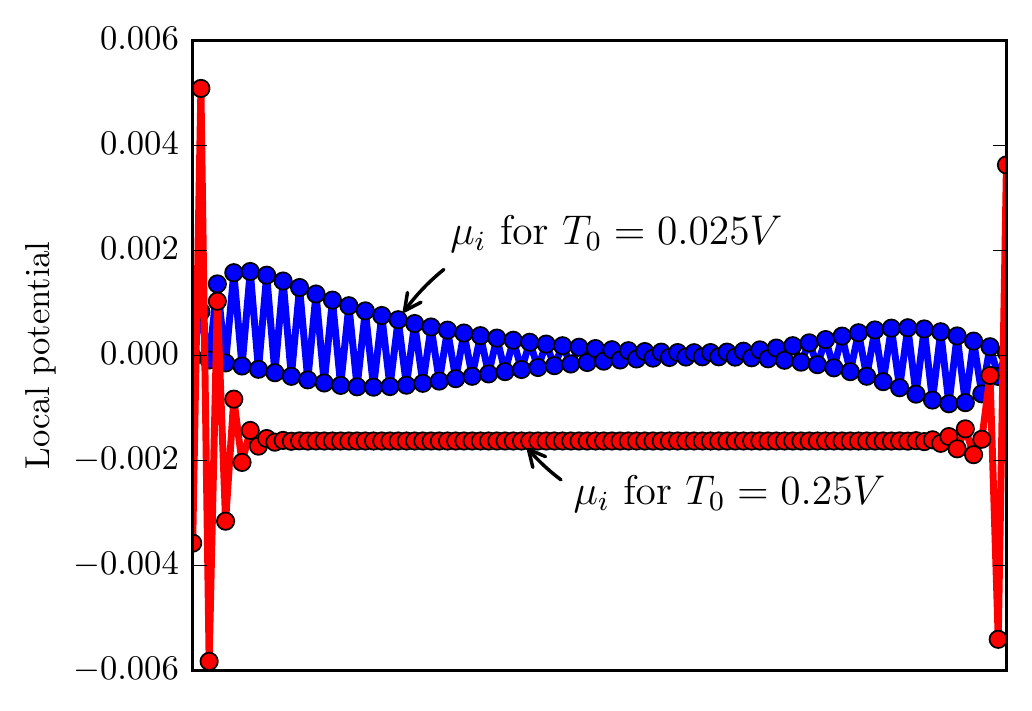}
    \includegraphics[width=.45\textwidth,clip]{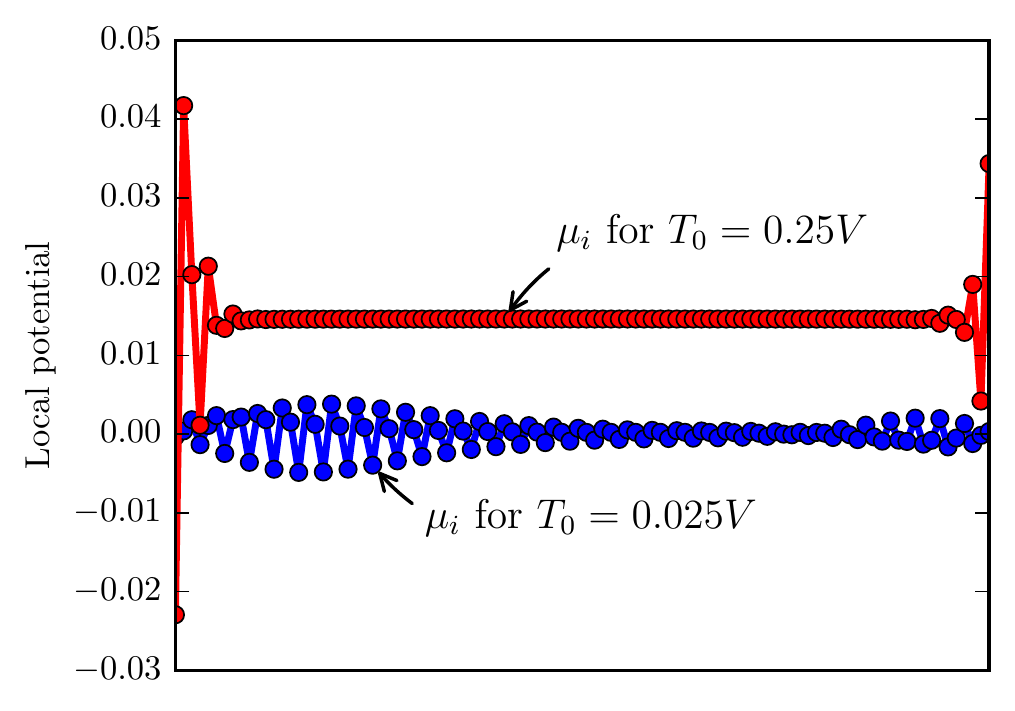}
    \caption{(color online) Plots of the local potential of the nanowire
    studied in Sec.\ \ref{SEC:Wire}. The upper panel depicts the local potentials for the
    wire without an external gate and the lower panel with gate. Similar to the local
    temperature shown in Fig.\ref{FIG:NanoWireLocalTemperature} the local potential
    exhibits Friedel oscillations with wave vector $q = 2 k_\mathrm{F}$ for a low initial temperature.
    \label{FIG:NanoWireLocalPotential}}
  \end{figure}

  In this appendix we provide additional plots. 
  In Fig.\ \ref{FIG:NanoWire3} and \ref{FIG:NanoWire4} we show snapshots of the spatial
  profiles of the transient density and energy wave at low temperatures. 
  In Fig.\ \ref{FIG:NanoWireLocalPotential} we show the local potential determined from the
  steady-state density and energy of the nanowire.

  \section{Numerical details} \label{APP:Numerics}
    The numerical computation of the time-dependent observables use two
    facts: 1) The system is noninteracting which allows for a direct solution
    of the equations of motion for the field operators. 2) 
    The time evolution is triggered by a sudden change in the Hamiltonian. This means
    that we do not have to worry about time-ordering.
    The main complication comes due to the ``openness'' of the system, i.e., the
    coupling of a finite system to semi-infinite leads.
    It has be shown recently that if the leads are treated in the wide-band limit,
    the time-evolution can solved almost analytically.\cite{TuovinenStefanucci:13,TuovinenVanLeeuwen:14}
    In our calculation we do not take the wide-band limit and therefore we
    have to rely on a numerically solution of the involved integrals.
    In the following we provide a rough sketch of the numerical implementation,
    focusing on two key aspects: The evaluation of the Matsubara summation
    needed to represent the initial state, and the technique to compute the
    Fourier transform leading to the single-particle propagators. An introduction
    to nonequilibrium quantum systems may be found in Ref.\ \onlinecite{StefanucciVanLeeuwen:13}.
    
    Since the Hamiltonian \eqref{Hamiltonian}, given in Sec.\ \ref{SEC:ThermoelectricTransport},
    is noninteracting, we can formally solve for the time-dependent fields operators ($\hbar=1$):
    \begin{subequations} \label{tdFields}
      \begin{align}
        \vec{\FF}(t) & = \dwi{\omega} e^{-i \omega t} \mat{G}^\Ret(\omega) \cdot \nn
        & \phantom{=} {} \lP \vec{\FF} 
        + \sum_{\alpha k} \Vakc  g^\Ret_{\alpha k}(\omega) \FF_{\alpha k} \rP ~, \label{FF} \\
        \FF_{\alpha k}(t) & = \dwi{\omega} e^{-i \omega t} \Bigg(
        g^\Ret_{\alpha k}(\omega) \FF_{\alpha k}
        + g^\Ret_{\alpha k}(\omega) \Vak \cdot \mat{G}^\Ret(\omega) \cdot \nn
        & \phantom{=} {} \lB \vec{\FF} + \sum_{\alpha' k'} \Vakcp
        g^\Ret_{\alpha' k'}(\omega) \FF_{\alpha' k'} \rB \Bigg) ~, \label{FFak}
      \end{align}
    \end{subequations}
    where $\vec{\FF}$ denotes the vector of field operators referring to the
    central region.
    In Eq.\ \eqref{tdFields} we have introduced the device Green's function
    \begin{align}
      \mat{G}(z) = \lP z - \mat{H} - \mat{\Sigma}(z) \rP^{-1} ~, \label{GFz}
    \end{align}
    given in terms of the embedding self-energy,
    \begin{align}
      \mat{\Sigma}(z) & = \sum_{\alpha k} \Vakc g_{\alpha k} (z) \Vak ~, \label{Sigmaz}
    \end{align}
    and the Hamiltonian of the central region.
    $\mat{\Sigma}(z)$, in turn, is given in terms of the bare Green's functions of the leads,
    \begin{align}
      g_{\alpha k}(z) & = \lP z - \eak \rP^{-1} ~. \label{gakz}
    \end{align}
    
    Using the explicit solution for the field operators we can write the time-dependent
    observables in terms of the initial state density matrices for the central region,
    \begin{align}
      \lA \vec{\FFd} \vec{\FF} \rA & = \dwi{\epsilon} f(\epsilon) 
      \lB \mat{G}^\Adv(\epsilon) - \mat{G}^\Ret(\epsilon) \rB ~, \label{deviceRDM}
    \end{align}
    the boundary of the central region and the leads,
    \begin{subequations}
      \begin{align}
        \lA \vec{\FFd} \FF_{\alpha k} \rA & = \dwi{\epsilon} f(\epsilon)
        \Big[ g^\Adv_{\alpha k}(\epsilon) \Vak \cdot \mat{G}^\Adv(\epsilon) \nn
        & \phantom{=} {} - g^\Ret_{\alpha k}(\epsilon) \Vak \cdot \mat{G}^\Ret(\epsilon) \Big] ~, \label{surfaceRDM1} \\
        \lA \FFd_{\alpha k} \vec{\FF} \rA & = \dwi{\epsilon} f(\epsilon)
        \Big[ \mat{G}^\Adv(\epsilon) \cdot \Vakc g^\Adv_{\alpha k}(\epsilon) \nn
        & \phantom{=} {} - \mat{G}^\Ret(\epsilon) \cdot \Vakc g^\Ret_{\alpha k}(\epsilon) \Big] ~, \label{surfaceRDM2}
      \end{align}
    \end{subequations}
    and the leads,
    \begin{align}
      \lA \FFd_{\alpha' k'} \FF_{\alpha k} \rA & = \dwi{\epsilon} f(\epsilon)
      \Bigg( \delta_{\alpha \alpha'} \delta_{k k'} \lB g^\Adv_{\alpha k}(\epsilon)
      - g^\Ret_{\alpha k}(\epsilon) \rB \phantom{ \mat{G}^\Adv(\epsilon) }  \nn
      & {} + \Big[ g^\Adv_{\alpha k}(\epsilon) \Vak \cdot \mat{G}^\Adv(\epsilon) \cdot \Vakcp
      g^\Adv_{\alpha' k'}(\epsilon) \label{leadsRDM} \\
      & {} - g^\Ret_{\alpha k}(\epsilon) \Vak \cdot \mat{G}^\Ret(\epsilon) \cdot \Vakcp
      g^\Ret_{\alpha' k'}(\epsilon) \Big] \Bigg) ~. \nonumber
    \end{align}
    
    In order to numerically evaluate integrals of the form
    \begin{align}
      \lA \FFd_\beta \FF_\alpha \rA = \dwi{\epsilon} f(\epsilon) 
      \lB G^\Adv_{\alpha \beta}(\epsilon) - G^\Ret_{\alpha \beta}(\epsilon) \rB ~, \label{InitialRDM3}
    \end{align}
    we use the following representation of the Fermi function:
    \begin{align}
      f(z) = \frac{1}{2} - \sum_f \frac{R_f}{z - i z_f} ~. \label{FermiFunctionRepresentation}
    \end{align}
    The residues $R_f$ and the \emph{modified} Matsubara frequencies $z_f$ can be obtained from the matrix
    \begin{align}
      B_{j j+1} = B_{j+1 j} = \frac{1}{2\sqrt{(2 j + 1) (2 j + 3)}} \;\; , \;\; 0 \leq j ~. \label{Bdefinition}
    \end{align}
    Considering the eigenvalue problem 
    \begin{align}
      \mat{B} \cdot \vec{b}_f = b_f \vec{b} ~, \label{BeigenSystem} 
    \end{align}
    it can be shown\cite{Ozaki:07} that $z_f$ and $R_f$ are given by
    \begin{subequations} \label{modifiedMatsubaraFrequencies}
      \begin{align}
        z_f & = \frac{1}{\beta b_f} ~, \label{zf} \\
        R_f & = \frac{1}{\beta} \lP \frac{\vec{b}_{f,0}}{2 b_f} \rP^2 ~, 
      \end{align}
    \end{subequations}
    where $\vec{b}_{f, 0}$ denotes the component $j=0$ of the eigenvector. Now we can use
    Eq.\ \eqref{FermiFunctionRepresentation} in Eq.\ \eqref{InitialRDM3} to obtain
    \begin{align}
      \lA \FFd_\beta \FF_\alpha \rA & = \sum_f R_f G^\Mat_{\alpha \beta}(i z_f) \nn
      & \phantom{=} {} + \ahalf \dwi{\epsilon} \lB G^\Adv_{\alpha \beta}(\epsilon)
      - G^\Ret_{\alpha \beta}(\epsilon) \rB ~. \label{InitialRDM4}
    \end{align}
    It has been shown that the truncated summation over $z_f$ converges much faster than
    the original Matsubara summation.\cite{KarraschSchoenhammer:10}

    \begin{figure}
      \includegraphics[width=.45\textwidth,clip]{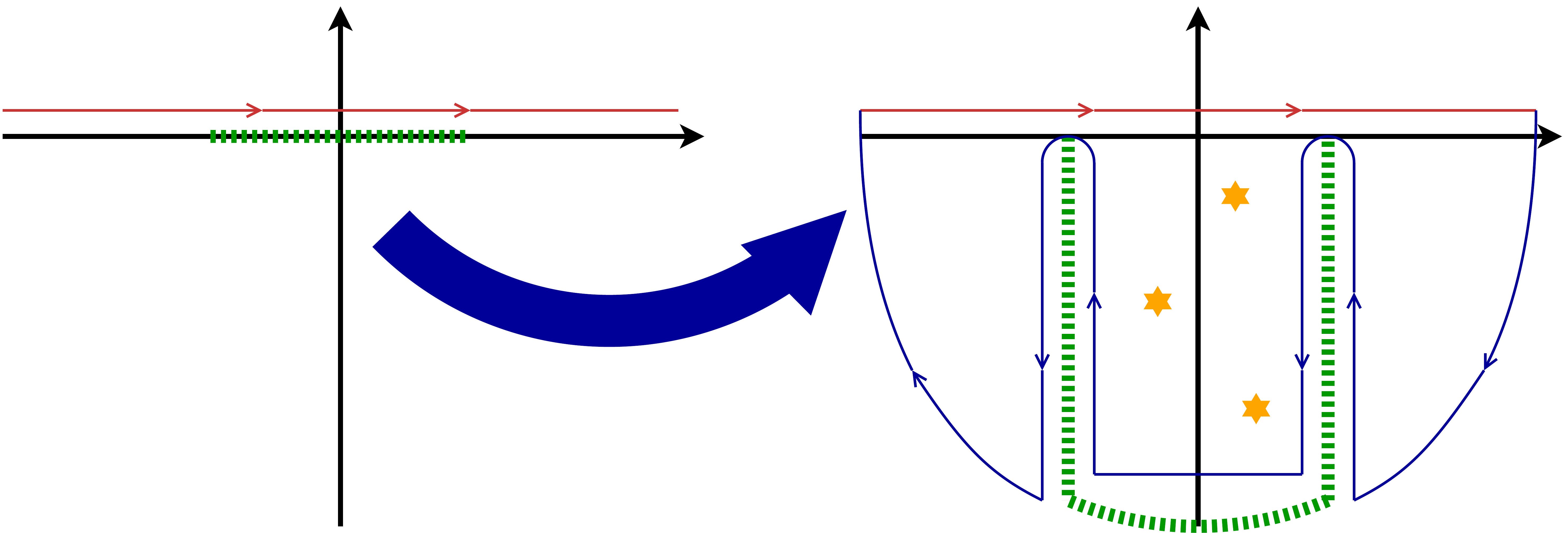}
      \caption{(color online) Sketch showing how the original Fourier integration
      contour (red) for a retarded integrand is replaced by an interrupted semi circle (blue)
      in the lower half of the complex plane. However, in order to do so the branch cut (green dashes)
      needs to be rotated away from the real axis and oriented along the negative imaginary axis.
      In deforming the branch cut we analytically continue the retarded function into the
      lower half of the complex plane, which potentially ``uncovers'' poles (orange stars),
      e.g., poles representing the quasi-particle energy for the case of the Green's function.
      The contribution from the contour on the
      semi circle vanish due to the Fourier exponential. The remaining contour, running
      back and forth along the branch cut is denoted by $\Cw$ in Eq.\ \eqref{FourierIntegral}.
      \label{FIG:branchCut}}
    \end{figure}
    For the calculation of the propagators we have to perform
    Fourier integrals of the type
    \begin{align}
      \dw{\omega} e^{\mp i \omega t} F^{\Ret/\Adv}(\omega) ~. \label{FourierTransform}
    \end{align}
    A straight-forward numerical evaluation is hampered by 
    a strongly oscillating integrand for $t \gg \tau$, where $\tau$ is a characteristic
    time scale of the Hamiltonian. This can be avoided by closing the integration
    contour with an infinite semi-arc in the lower/upper half of the complex plane for
    $F^\Ret(\omega)$/$F^\Adv(\omega)$. However, the function $F^{\Ret/\Adv}$ may has branch cuts
    on the real axis due to the embedding self-energy. Figure \ref{FIG:branchCut}
    shows how the branch cut can be rotated away from the real axis and directed along the negative
    (or positive) imaginary axis.
    The semi-arc has to be interrupted with integration contours running along the deformed
    branch cuts. We label these contours by $\Cw$. This allows us to write the Fourier transform as
    \begin{align}
      & \dw{\omega} e^{\mp i \omega t} F^{\Ret/\Adv}(\omega) = 
      - \Bigg( \dz{\omega}{\Cw} e^{\mp i \omega t} F^{\Ret/\Adv}(\omega) \nn
      & \pm i \sum_m e^{\mp i \omega_m t} \Res{F^{\Ret/\Adv}(\omega_m)} \Bigg) ~, \label{FourierIntegral} 
    \end{align}
    where $\omega_m$ are the poles in the lower/upper half of the complex plane of $F^{\Ret/\Adv}(z)$,
    respectively. Since the contour $\Cw$ is always parallel to the imaginary axis, the Fourier
    exponentials are now exponentially decaying, which improves the numerical stability and allows us to
    compute the long-time behavior accurately and efficiently.
    
\end{appendix}

\bibliography{ThermalTransients}

\end{document}